\documentclass[reprint,amsmath,amssymb,aps,prl,shortbibliography,]{revtex4-2}

\usepackage{graphicx}
\usepackage{amsmath}
\usepackage{xcolor}
\usepackage{dcolumn}
\usepackage{enumitem}
\usepackage{epstopdf}

\begin{document}

\title{Approaching to the deep-strong photon-to-magnon coupling}

\author{I.~A.~Golovchanskiy$^{1,2}$, N.~N.~Abramov$^{2}$, V.~S.~Stolyarov$^{1,3}$, A.~A.~Golubov$^{1,4}$, M.~Yu.~Kupriyanov$^{1,5}$, V.~V.~Ryazanov$^{2,6}$, A.~V.~Ustinov$^{2,7}$}

\affiliation{
$^{1}$ Moscow Institute of Physics and Technology, State University, 9 Institutskiy per., Dolgoprudny, Moscow Region, 141700, Russia; \\
$^{2}$ National University of Science and Technology MISIS, 4 Leninsky prosp., Moscow, 119049, Russia; \\
$^{3}$ Dukhov Research Institute of Automatics (VNIIA), Sushchevskaya 22, Moscow 127055, Russia; \\
$^{4}$ Faculty of Science and Technology and MESA+ Institute for Nanotechnology,
University of Twente, 7500 AE Enschede, The Netherlands; \\
$^{5}$ Skobeltsyn Institute of Nuclear Physics, MSU, Moscow, 119991, Russia \\
$^{6}$ Institute of Solid State Physics (ISSP RAS), Chernogolovka, 142432, Moscow region, Russia; \\
$^{7}$ Physikalisches Institut, Karlsruhe Institute of Technology, 76131 Karlsruhe, Germany.
}%

\begin{abstract}
In this work, the ultra-strong photon-to-magnon coupling is demonstrated for on-chip multilayered superconductor/ferromagnet/insulator hybrid thin film structures
reaching the coupling strength above 6~GHz, the coupling ratio about 0.6, the single-spin coupling strength about 350~Hz, and cooperativity about $10^4$.
High characteristics of coupling are achieved owing to a radical suppression of the photon phase velocity in electromagnetic resonator.
With achieved coupling the spectrum reveals inapplicability of the Dicke quantum model, and evidences contribution of the diamagnetic $A^2$ interaction term in the Hamiltonian of the system, which satisfies the Thomas-Reiche-Kuhn sum rule.
The contribution of the $A^2$ term denotes validity of the Hopfield quantum model and manifests observation of a different hybrid polariton quasi-particle, namely, the plasmon-magnon polariton.
\end{abstract}

\maketitle


Concepts of light-matter interaction are at the heart of modern quantum technologies \cite{Kockum_NatPhysRev_1_19,RMP_91_025005}. 
Irregardless the particular platform key characteristics of the light-matter interaction are the coupling strength $g$ and the coupling ratio $g/\omega$, where $\omega$ is the transition frequencies in the system.
When the coupling strength is increased beyond the so-called weak and strong coupling regimes and reaches a considerable fraction of the transition frequencies in the system the ultra-strong ($g/\omega>0.1$) or the deep-strong ($g/\omega\gtrsim 1$) coupling regime is reached.
In these regimes the standard quantum optical approximations, the rotating wave approximation (RWA), fail \cite{Kockum_NatPhysRev_1_19,RMP_91_025005,Flower_NJP_21_095004}.
In ultra-strong coupling regime it starts to become possible to modify the nature of the light and matter degrees of freedom: 
the global vacuum energy of the system becomes dependent on the coupling strength \cite{Ciuti_PRB_72_115303,Baranov_NatComm_11_2715};
the ground state gains a photonic component \cite{Ciuti_PRB_72_115303,Todorov_PRB_85_045304}.
In deep-strong coupling regime electromagnetic states fundamentally change the physical properties of a matter \cite{Mueller_Nat_583_780,NatPhys_13_44,NanoLett_17_6340}.
Conventionally, the light-matter interaction is refereed to via polariton hybrid quasi-particles, e.g., phonon polariton, plasmon polariton, etc. 

Quantum magnonics \cite{Tabuchi_CRR_17_729,Lachance-Quirion_APE_12_070101,Huebl_PRL_111_127003,Tabuchi_PRL_113_083603,Zhang_PRL_113_156401,Rameshti_PRB_91_214430,Flower_NJP_21_095004,Li_PRL_123_107701,Hou_PRL_123_107702,Golovchanskiy_Sci} is one of emerging fields in hybrid quantum technologies \cite{Xiang_RMP_85_623,Clerk_NatPhys_16_257}, 
which considers a coherent coupling of collective spin excitations with photons. 
Quantum magnonics offers novel approaches, including hybrid magnonic-based quantum platforms \cite{Tabuchi_Sci_349_405,Lachance-Quirion_Sci_367_425,Wang_arXiv}, magnon memory \cite{Zhang_NatComm_6_8914}, and microwave-to-optical quantum transducers \cite{Hisatomi_PRB_93_174427}.
It is anticipated \cite{Flower_NJP_21_095004} that quantum magnonic hybrids obey the Dicke model \cite{Kockum_NatPhysRev_1_19,Kirton_AdvQT_2_1800043}.

The major restriction for developments in quantum magnonics is imposed by a weak coupling strength between photons and magnons.
Originally, the strong photon-to-magnon coupling regime \cite{Huebl_PRL_111_127003,Tabuchi_PRL_113_083603,Zhang_PRL_113_156401}, and later the ultra-strong photon-to-magnon coupling \cite{Rameshti_PRB_91_214430,Flower_NJP_21_095004,Bourhill_PRB_93_144420} were achieved in truly macroscopic systems (dimensions reach several millimeters) by utilizing the Dicke coupling relation \cite{Huebl_PRL_111_127003,Tabuchi_PRL_113_083603, Zhang_PRL_113_156401,Lachance-Quirion_APE_12_070101,Kirton_AdvQT_2_1800043} $g=g_s\sqrt{N}$, where $g_s$ is the single spin coupling strength and $N$ is the number of spins in the system.
For practical realization on-chip quantum systems are highly desired, which requires high single-spin coupling strength $g_s$. 
Only recently the on-chip strong \cite{Li_PRL_123_107701,Hou_PRL_123_107702} and ultra-strong \cite{Golovchanskiy_Sci} photon-to-magnon coupling regimes were demonstrated. 
Though, demonstrated systems suffer from large relaxation rates manifested in poor cooperativity \cite{Li_PRL_123_107701,Hou_PRL_123_107702,Golovchanskiy_Sci}.
Also, demonstrated coupling still is insufficient for verification of the interaction model, as the Dicke model is subjected to the super-radiant phase transition.

In this work, the ultra-strong photon-to-magnon coupling is demonstrated in on-chip thin film hybrid structures with the coupling strength exceeding $g/2\pi\gtrsim 6$~GHz, the coupling ratio $g/\omega\approx0.6$, the single-spin coupling strength $g_s\approx 350$~Hz, and cooperativity about $10^4$.
High characteristics of coupling are achieved with on-chip multilayered hybrid film structures 
owing to a radical suppression of the photon phase velocity in electromagnetic resonator.
With achieved coupling ratio $g/\omega\approx0.6\sim 1$ the system approaches to the deep-strong coupling regime. 
At this coupling regime it is verified that the Dicke model is inapplicable, the super-radiant does not take place.
Instead, the spectrum clearly evidences contribution of the diamagnetic $A^2$ interaction term in the Hamiltonian of the system, which satisfies the Thomas-Reiche-Kuhn sum rule.
The contribution of the $A^2$ term verifies validity of the most general Hopfield light-matter interaction model \cite{Kockum_NatPhysRev_1_19,Mueller_Nat_583_780,Baranov_NatComm_11_2715,Hopfield_PR_112_1555} and manifests observation of a different hybrid polariton quasi-particle, namely, the plasmon-magnon polariton.


%
\begin{figure}[!ht]
\begin{center}
\includegraphics[width=0.7\columnwidth]{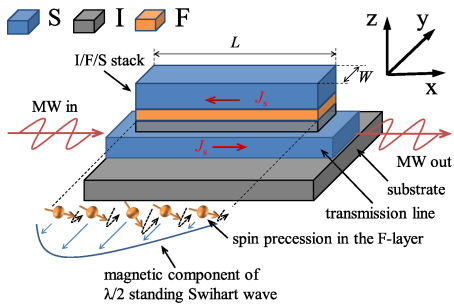}
\caption{
Schematic illustration of the investigated chip-sample.  
A series of insulator/ferromagnet/superconductor (I/F/S) film rectangles is placed directly on top of the central transmission line of superconducting (S) waveguide.
Magnetic field $H$ is applied in-plane along the $x$-axis.
MW in/out indicates the microwave transmission.
Blue curve and blue arrows indicate the magnetic field component of the half-lambda standing electromagnetic Swihart wave, which is induced by superconducting sheet currents $J_s$ in Nb layers oscillating along the x-direction (red arrows).
Interaction of the Swihart standing wave with spins in ferromagnetic layer (indicated with orange arrows) result in level repulsion of resonance lines.
}
\label{sam}
\end{center}
\end{figure}

A schematic illustration of investigated system is shown in Fig.~\ref{sam}.
The system consists of superconducting (niobium) film microwave transmission line of thickness 300~nm with multilayered insulator/ferromagnet/superconductor (I/F/S=silicon/permalloy/niobuim) rectangular film heterostructures of length $L=1.1$~mm, width 130~$\mu$m and thickness 15~nm/25~nm/230~nm, respectively, placed directly on top of the transmission line.
In general, such multilayer should be viewed as a combination of two interacting subsystems \cite{Golovchanskiy_Sci}.
The first subsystem is the electromagnetic resonator that is formed between two superconducting layers separated by the insulator.
The electromagnetic resonator has suppressed Swihart photon phase velocity $\overline{c}=c_0\sqrt{d_I/\varepsilon_I(2\lambda_L+d_I+d_F)}$, where $c_0$ is the velocity of light, $d_{I(F)}$ is the thickness of the insulating (ferromagnetic) layer, $\varepsilon_I$ is the dielectric constant of the I-layer and $\lambda_L$ is the London penetration depth of S-layers.
Considering $\varepsilon_I\approx10$ for Si, and $\lambda_L\approx80$~nm in Nb the obtained phase velocity in the electromagnetic resonator $\overline{c}\approx0.007c_0$ provides the estimation for the Swihart resonance frequency $\overline{c}/2L\approx9.7$~GHz for $\lambda/2$ standing wave resonance.
The second subsystem is the conventional ferromagnetic permalloy film placed inside the microwave resonator.
Microwave response of the sample is studied by analyzing the field-derivative of the transmitted microwave signal $dS_{21}(f,H)/dH$ obtained with the vector network analyzer (VNA) at various fields and temperatures.
More details of sample fabrication and measurements are given in supplementary materials. 

\begin{figure*}[!ht]
\begin{center}
\includegraphics[width=0.48\columnwidth]{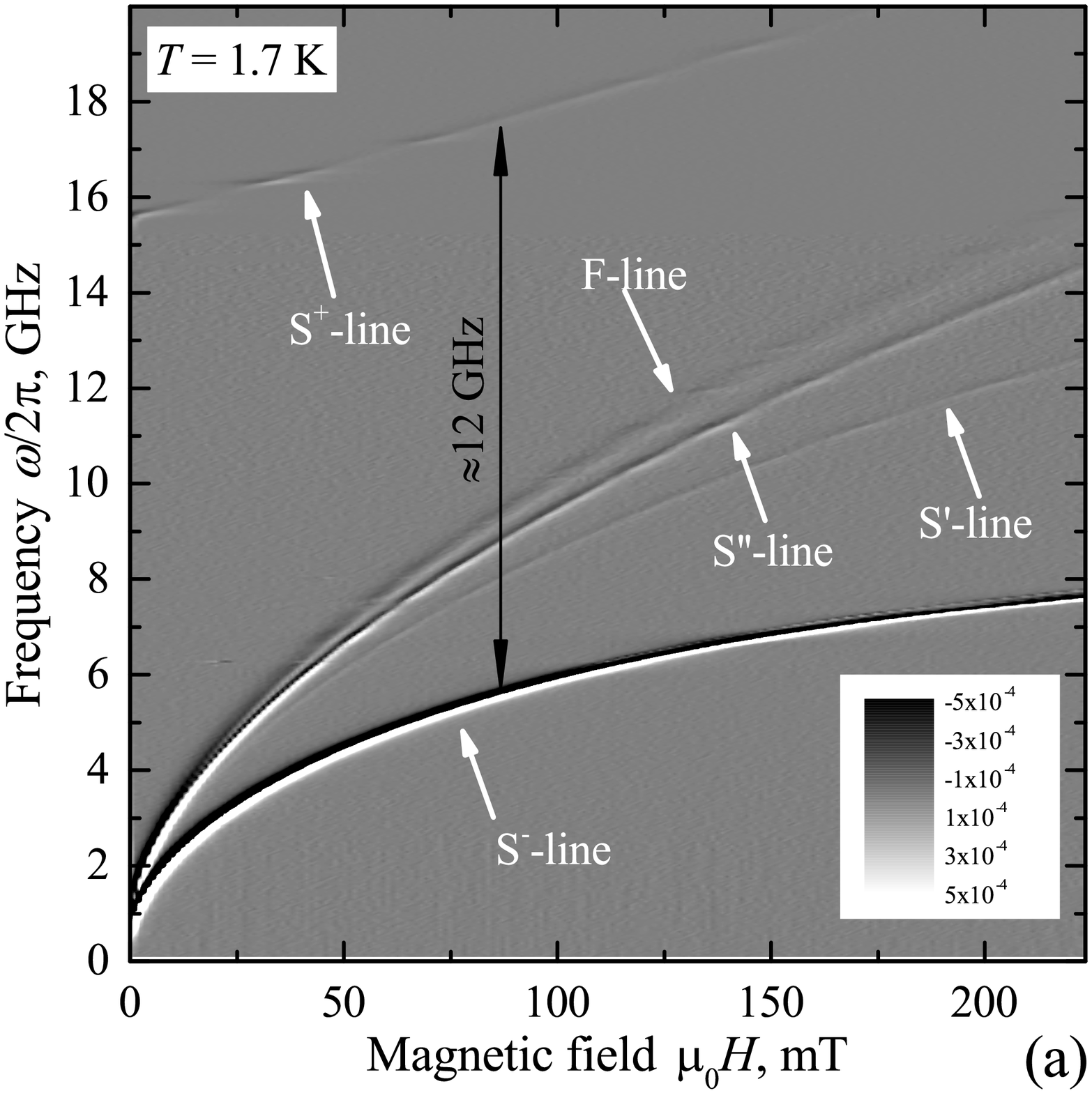}
\includegraphics[width=0.48\columnwidth]{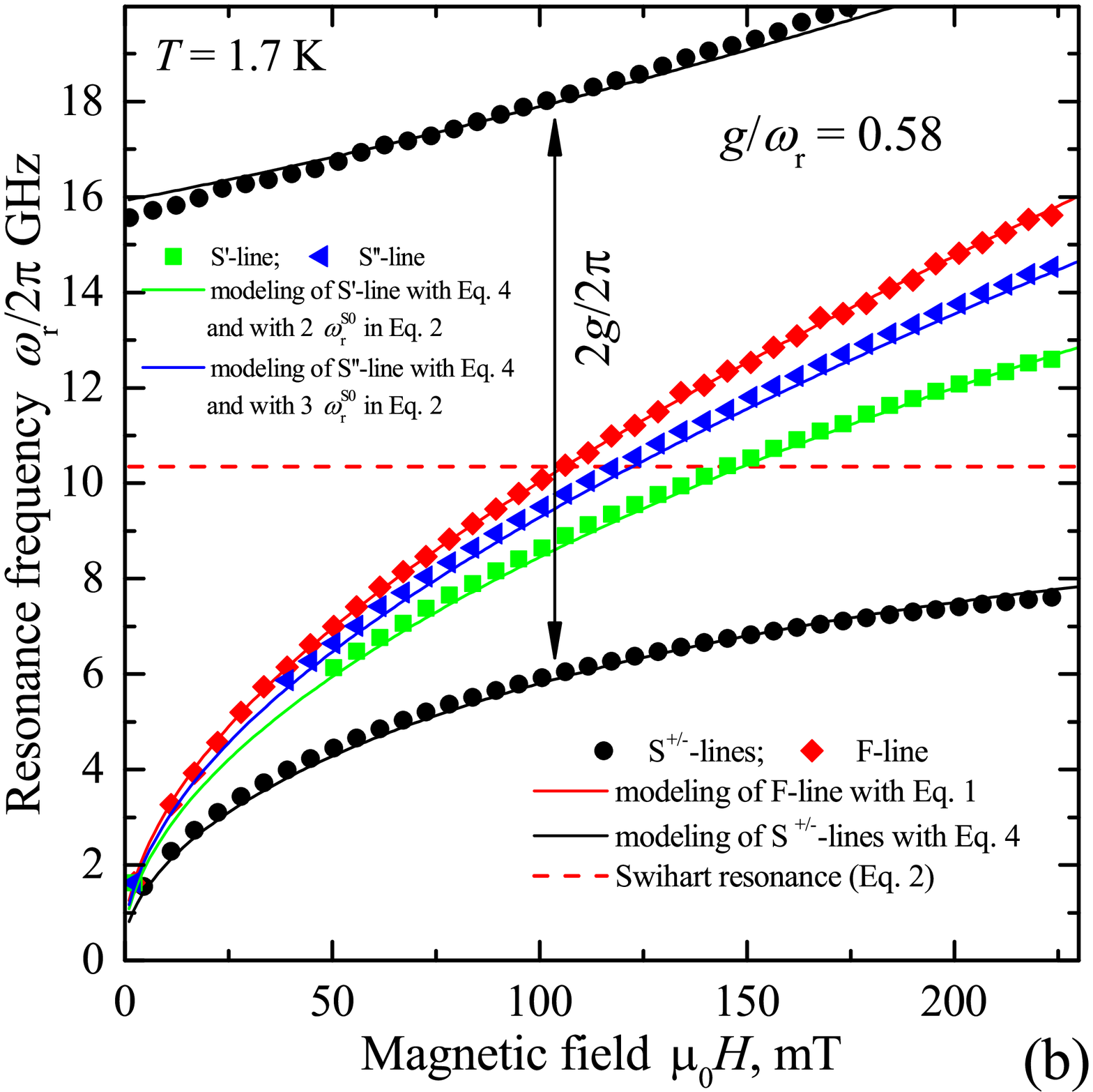}
\includegraphics[width=0.48\columnwidth]{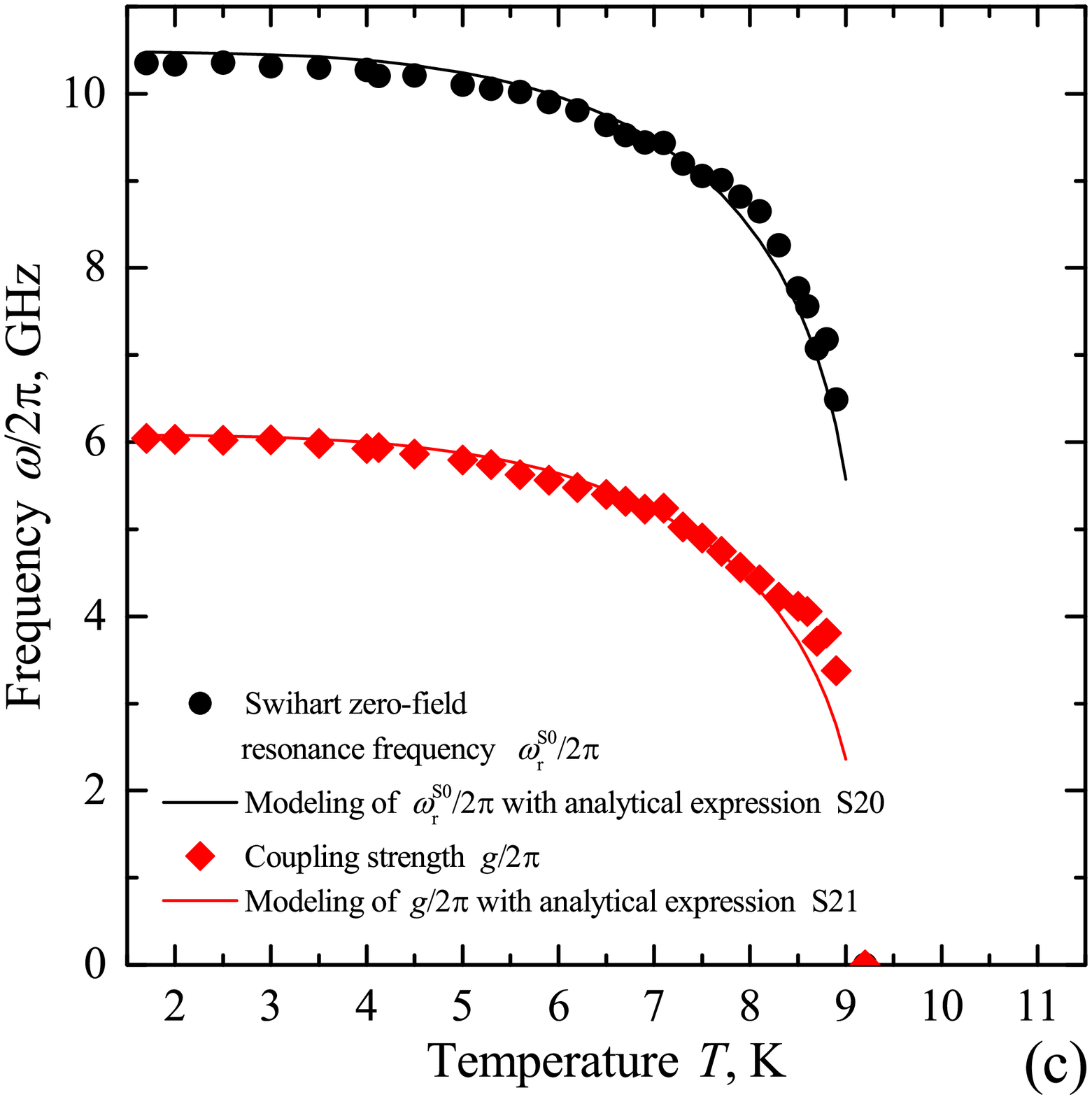}
\includegraphics[width=0.538\columnwidth]{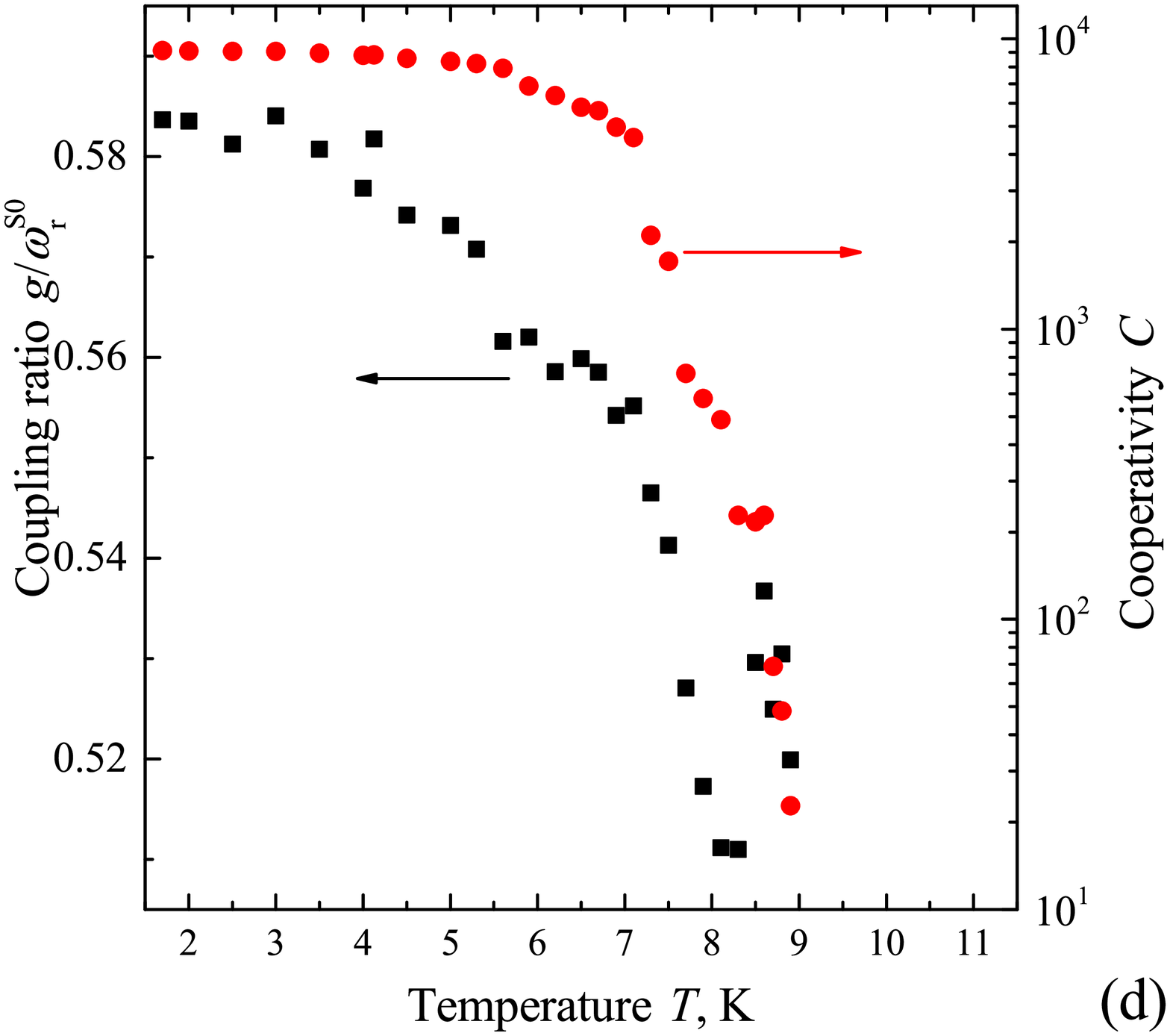}
\caption{
a) Microwave transmission spectrum $dS_{21}(f,H)/dH$ of the studied sample measured at temperature below (b) the superconducting critical temperature.
Black arrow in (b) indicates the level-repulsion of about 12 GHz.
b) Modeling of spectral lines at $T=1.7$~K with Eqs.~\ref{FMR}, \ref{Swi}, and \ref{Freq}.
Experimental F-, S$^{+/-}$-lines are shown with red and black symbols.
The optimum fit for the F-line with Eq.~\ref{FMR} yields $\mu_0 H_a=0.7$~mT and $\mu_0 M_{eff}=1.05$~T (solid red curve).
The optimum fit for S$^+$- and S$^-$ lines with Eqs.~\ref{FMR}, \ref{Swi}, and \ref{Freq} yields $\omega^{S0}_r/2\pi=10.35$~GHz, the parameter $\alpha\sim10^{-14}$~T$^{-2}$, and $g/2\pi=6.04$~GHz.
Dashed red curve shows  $\omega^{S}_r(H)$ (Eq.~\ref{Swi}).   
Blue and green symbols show experimental S'- and S''-lines, respectively.
Blue and green curves are obtained as S$^-$-lines in Eq.~\ref{Freq} when the $\omega^{S0}_r$ in Eq.~\ref{Swi} is substituted with $2\omega^{S0}_r$ and $3\omega^{S0}_r$, respectively.
c) Dependencies of the coupling strength on temperature $g(T)/2\pi$, and of the Swihart resonance frequency on temperature $\omega^{S0}_r(T)/2\pi$.
Solid lines in (b) show modeling of $g(T)/2\pi$ and $\omega^{S0}_r(T)/2\pi$ with analytical expressions, which yield $g_0/2\pi=6.1$~GHz, and $\omega^{S0}_r(0)/2\pi=10.48$~GHz.
d) The dependence of the coupling ratio $g/\omega^{S0}_r(T)$ and of the cooperativity $C(T)$ on temperature.
}
\label{Exp1}
\end{center}
\end{figure*}

Figure~\ref{Exp1}a show transmission spectra of the sample at temperature well below the superconducting critical temperature of Nb ($T_c \approx 9.1$~K). 
The spectrum consist of five resonance lines which can be identified as follows \cite{Golovchanskiy_Sci}:
(i) S$^+$- and S$^-$-lines are the level-repulsion resonance lines, which represent coherent polaritonic interaction between the electromagnetic Swihart resonator and the ferromagnetic layer;
(ii) F-line indicates the ferromagnetic resonance (FMR), which emerges due to FMR absorption at edge areas of the permalloy rectangle \cite{Golovchanskiy_Sci}; 
(iii) additional S'- and S''-lines are signatures of higher-order oscillations and will be specified below.
Ultra-strong coupling between the electromagnetic and ferromagnetic resonators $g/2\pi$ is represented by large frequency split between repulsed S$^+$- and S$^-$-lines exceeding 10~GHz in comparison to estimated Swihart resonance frequency 9.7~GHz.
Ultra-strong coupling is achieved owing to (i) reduced magnetic volume $V_m$ of the Swihart electromagnetic resonator $g\propto 1/\sqrt{V_m}$ \cite{Huebl_PRL_111_127003,Zhang_PRL_113_156401,Li_PRL_123_107701,Hou_PRL_123_107702,Golovchanskiy_Sci} due to suppressed velocity of light $\overline{c}$, and owing to (ii) maximized amplitude of microwave magnetic field in the F-layer \cite{Golovchanskiy_Sci}. 
At $T>T_c$ all spectral lines except of the F-line vanish, the position of the F-line remains unchanged.

Individually, electromagnetic and ferromagnetic resonators follow known dependencies on magnetic field.
The FMR line (F-line) represent the magnon eigenfrequencies and follows the typical dependence of the FMR frequency of thin film on in-plane external magnetic field.
\begin{equation}
(\omega^{F}_r(H)/\mu_0\gamma)^2=H_{eff}(H_{eff}+M_{eff})
\label{FMR}
\end{equation}
where $\mu_0$ is the vacuum permeability, $\gamma$ is the gyromagnetic ratio, $H_{eff}=H+H_a$ is the in-plane effective field, which includes the in-plane anisotropy term $H_a$ along the applied field, and $M_{eff}$ is the effective magnetization field.
Modeling of the F-line with Eq.~\ref{FMR} at different temperatures yields magnetic parameters that are typical for permalloy: 
the anisotropy field $\mu_0 H_a\approx 1$~mT, the effective magnetization $\mu_0 M_{eff}\approx 1.05$~T, and no significant dependence of these parameters on temperature (see supplementary). 
The electromagnetic Swihart resonance $\omega^{S}_r$ represents the photon eigenfrequency and  follows the dependence on field \cite{Golovchanskiy_Sci} (see supplementary).
\begin{equation}
\omega^{S}_r(H)=\omega^{S0}_r/\sqrt{1+\alpha H^2}
\label{Swi}
\end{equation}
where $\alpha$ is a free parameter, and $\omega^{S0}_r/2\pi\approx 9.7$~GHz is the zero-field Swichard resonance frequency.

Quantitative description of the photon-to-magnon interaction requires to specify the interaction model.
The Hamiltonian $\hat{H}$ of any hybrid polariton system consists of three terms:
\begin{equation}
\hat{H}=\hat{H}_p+\hat{H}_m+\hat{H}_I
\label{Ham}
\end{equation}
where $\hat{H}_p\approx\omega_p\hat{a}^\dagger\hat{a}$ is the Hamiltonian of the photon subsystem, $\omega_p=\omega^{S}_r$ is the photon eigen-frequency, and $\hat{a}^\dagger$ ($\hat{a}$) are the photon creation (annihilation) operators;
$\hat{H}_m\approx\omega_m\hat{b}^\dagger\hat{b}$ is the Hamiltonian of the magnon subsystem, $\omega_m=\omega^{F}_r$ is the magnon eigen-frequency, and $\hat{b}^\dagger$ ($\hat{b}$) are the magnon creation (annihilation) operators;
and $\hat{H}_I$ is the interaction term.
In general, the interaction in bosonic systems is represented by the interaction term for coupled harmonic oscillators \cite{Flower_NJP_21_095004,Kirton_AdvQT_2_1800043} $\hat{H}_I=g(\hat{a}^\dagger+\hat{a})(\hat{b}^\dagger+\hat{b})$ (see supplementary).
With the specified coupling term and with coupling strength \cite{Huebl_PRL_111_127003,Tabuchi_PRL_113_083603, Zhang_PRL_113_156401,Lachance-Quirion_APE_12_070101,Kirton_AdvQT_2_1800043} $g=g_s\sqrt{N}$, Eq.~\ref{Ham} forms the Dicke Hamiltonian \cite{Kockum_NatPhysRev_1_19,Kirton_AdvQT_2_1800043}.
In case when $g/\omega_p<0.1$ the interaction term can be relaxed \cite{Huebl_PRL_111_127003,Tabuchi_PRL_113_083603,Zhang_PRL_113_156401} to $\hat{H}_I=g(\hat{a}^\dagger\hat{b}+\hat{b}^\dagger\hat{a})$ following the rotating-wave approximation, and the Dicke model relaxes to the Tavis-Cummings model \cite{Kockum_NatPhysRev_1_19}.
Eigen-frequencies of hybridized polariton quasi-particles can be obtained from the Hamiltonian in Eq.~\ref{Ham} by using the Hopfield-Bogolubov transformation \cite{Hopfield_PR_112_1555} (see supplementary).

However, the spectrum in Fig.~\ref{Exp1}a can not be analyzed employing the Dicke model directly.
With the estimated coupling strength $2g/2\pi\approx12$~GHz (see Fig.~\ref{Exp1}a) and the estimated Swihart resonance frequency $\omega^{S0}_r/2\pi\approx 9.7$~GHz the coupling ratio clearly exceeds 0.5.
At $g/\omega>0.5$ one polariton eigen-frequency becomes complex-valued.
This indicates that the Dicke system undergoes the super-radiant phase transition \cite{Kirton_AdvQT_2_1800043,Nataf_NatCom_1_72,Baumann_Nat_464_1301} where different Hamiltonian becomes applicable \cite{Emary_PRL_90_044101,Emary_PRE_67_066203}.
At classical super-radiant phase transition occurring at a finite temperature the system exhibits a spontaneous polarization of spins and a spontaneous coherent electromagnetic field \cite{Hepp_AoP_76_360}.
At quantum phase transition, at zero temperature, the vacuum (ground state) of the cavity system is twice degenerate. 
A linear superposition of these two degenerate ground states can be seen as a collective qubit characterized by a strong light-matter entanglement (the cavity field is in a coherent state, the matter part in a ``ferromagnetic'' phase) \cite{Baumann_Nat_464_1301}.

Potentially, occurrence of the super-radiant phase transition can be verified via observation of the so-called soft modes when the coupling ratio $g/\omega$ shows a dependence on some external parameter and crosses the transition point $g/\omega=0.5$. 
In this case, the lower-frequency polariton branch decreases anomalously as the transition point is reached, and than increases for the other super-radiant phase \cite{Nataf_NatCom_1_72,Emary_PRL_90_044101,Emary_PRE_67_066203}.
However, in this work the coupling ratio $g/\omega$ does not demonstrate required tunability.
On the other side, large coupling ratio allows to decide on appropriate coupling model by considering several separate cases as follows.

First, we consider that formally the super-radiant phase transition does occur (considered in details in supplementary). 
Upon the super-radiant transition bosonic modes are displaced as \cite{Emary_PRL_90_044101,Emary_PRE_67_066203} $\hat{a}^\dagger \rightarrow \hat{c}^\dagger +\sqrt{\alpha}$ and $\hat{b}^\dagger \rightarrow \hat{d}^\dagger -\sqrt{\beta}$, changing the Hamiltonian and the expression for transition frequencies.
The expression for transition frequencies shows a rather poor convergence with experimental frequencies and does not pass the confirmation.
Therefore, we confirm that the magnonic subsystem is not in super-radiant phase.

Next, we consider that the F-line does not represent eigen-frequencies of the ferromagnetic resonator, which is coupled to the Swihart resonator, and that the eigen-frequencies of the Swihart resonator are underestimated.
Indeed, underestimation of $d_I$ or overestimation of $\varepsilon_I$ leads to higher  $\omega^{S0}_r$.
Also, interaction of the electromagnetic resonance with higher-frequency ferromagnetic resonance may be considered, since the F-line is rather an artifact of the system \cite{Golovchanskiy_Sci}.
It can be assumed that the magnon frequency is increased in comparison to the F-line due to Meissner stray fields \cite{Golovchanskiy_AdvSci_6_1900435}, due to the proximity effect \cite{Golovchanskiy_PRAppl_14_024086}, the electromagnetic proximity effect \cite{Mironov_APL_113_022601,Volkov_PRB_99_144506}, the inverse proximity effect \cite{Bergeret_EPL_66_111,Dahir_PRB_100_134513}, due to excitation of the magnetostatic standing waves \cite{Golovchanskiy_AFM_28_1802375,Golovchanskiy_JAP_124_233903}, or of the perpendicular standing waves \cite{Kittel_PR_100_1295,Seavey_JAP_30_S227}.
Such hypothetical scenarios are worked out in supplementary; these scenarios are not confirmed.
Therefore, the Dicke model is not applicable.

Alternatively, it can be noted that cavity hybrid systems can be protected from the super-radiant phase transition by the additional interaction term \cite{Kockum_NatPhysRev_1_19,Nataf_NatCom_1_72}, known as the diamagnetic term or $A^2$ term $\propto D(\hat{a}^\dagger+\hat{a})^2$, where $\hat{A}\propto (\hat{a}^\dagger+\hat{a})$ is the vector potential.
Though, it should be noted that so far the $A^2$ term was considered for electrostatically coupled systems only \cite{Nataf_NatCom_1_72,Mueller_Nat_583_780,Baranov_NatComm_11_2715,Hopfield_PR_112_1555}.
In this case, the complete form of the Hamiltonian interaction term is
\begin{equation}
\hat{H}_I=g\sqrt{\frac{\omega_m}{\omega_p}}(\hat{a}^\dagger+\hat{a})(\hat{b}^\dagger+\hat{b})+D(\hat{a}^\dagger+\hat{a})^2
\label{Int}
\end{equation}
where $D$ is the diamagnetic coupling factor.
In case when light-matter interaction is solely mediated by dipole interactions  the diamagnetic coupling factor can be expressed with the Thomas-Reiche-Kuhn sum rule \cite{Kockum_NatPhysRev_1_19} $D=g^2/\omega_m$.
Diagonalization of the Hamiltonian in Eq.~\ref{Ham} with the interaction term in Eq.~\ref{Int} yields the following bi-quadratic equation for spectral frequencies
\begin{equation}
{\omega^\pm}^4-{\omega^\pm}^2\left({\omega_p}^2+{\omega_m}^2+4g^2\right)+{\omega_p}^2{\omega_m}^2=0
\label{Freq}
\end{equation}

Equations~\ref{FMR},~\ref{Swi}, and \ref{Freq} are employed for modeling of spectral lines using the following routine.
First, the F-line is modeled separately with Eq.~\ref{FMR}.
Next, S$^{+/-}$-lines are fitted with Eq.~\ref{Freq} using $H_a$ and $M_{eff}$ as
fixed parameters and using parameters of the Swihart resonator (Eq.~\ref{Swi}) and the coupling strength $g$ as fitting parameters.
The result of modeling of the spectrum at $T=1.7$~K is shown in Fig.~\ref{Exp1}b. 
The optimum fit yields the zero-field Swihart resonance frequency $\omega^{S0}_r/2\pi=10.35$~GHz, and the coupling strength $g/2\pi=6.04$~GHz.

Validity of the interaction term in Eq.~\ref{Int} can be confirmed by considering the S'- and S''-lines.
Typically, a spectrum of a hybrid system that consist of two coupled harmonic oscillators contains only two polariton branches.
However, in case of sufficiently strong coupling the spectrum can incorporate additional lines in the range between the anti-crossing lines that appear as a result of hybridization of higher-order photon or magnon modes \cite{Zhang_PRL_113_156401,Rameshti_PRB_91_214430,Flower_NJP_21_095004}.
According to the dispersion for magnetostatic waves in superconductor/ferromagnet multilayers \cite{Golovchanskiy_AFM_28_1802375,Golovchanskiy_JAP_124_233903} higher-order magnon modes do not differ in frequency from the F-line due to small $kd_F$ product, where $k$ is the wavenumber of the standing magnetostatic wave.
Therefore, S'- and S''-lines reflect interaction of magnons with higher-order photon modes.
The latter are the Swihart standing wave resonances with the wavelength $n\lambda/2=L$, where $n$ is integer.
Importantly, the coupling strength $g$ is expected to be unchanged \cite{Golovchanskiy_Sci} since for higher-order photon modes the thin-film geometry is preserved, and, therefore, the  single-spin coupling strength is also preserved .
Green and blue curves in Fig.~\ref{Exp1}b show $\omega^-$ curves (Eq.~\ref{Freq}) when $\omega^{S0}_r$ in Eq.~\ref{Swi} is substituted with $n\omega^{S0}_r$ for $n=2$ and $n=3$ while the rest of parameters in Eqs.~\ref{FMR},~\ref{Swi}, and \ref{Freq} is unchanged.
A remarkable coincidence of the model curves with experimental S'- and S''-lines verifies the interaction term in Eq.~\ref{Int}.
In addition, it explains why S''-line in Fig.~\ref{Exp1}a is substantially stronger than the S'-line:
S''-line represents the coupling with the even $2\lambda/2$ electromagnetic mode, which should be coupled considerably weaker to the external microwave field in comparison to the odd $3\lambda/2$ mode, represented by S'-line, owing to its symmetry.

Transmission spectra of the sample has been measured and modeled for the temperature range from 1.7~K up to 11~K. 
Figure~\ref{Exp1}c shows the obtained temperature dependencies of the coupling strength $g(T)/2\pi$ and of the zero-field Swihard resonance frequency $\omega^{S0}_r(T)/2\pi$.
Modeling of $g(T)$ with analytical expression $g(T)=g_0(1-(T/T_c)^4)^{3/8}$ (see supplementary) yields the maximum coupling strength $g_0/2\pi=6.08$~GHz at zero temperature.
The zero-temperature coupling strength $g_0$ provides the single-spin coupling strength \cite{Huebl_PRL_111_127003,Tabuchi_PRL_113_083603, Zhang_PRL_113_156401,Lachance-Quirion_APE_12_070101,Kirton_AdvQT_2_1800043} $g_s/2\pi=g/2\pi/\sqrt{N}\approx350$~Hz, where $N=3.1\times10^{14}$ is the number of spins in the system.
Currently, this is the highest value for the single-spin coupling strength ever reported.
Modeling of $\omega^{S0}_r(T)$ with analytical expression $\omega^{S0}_r(T)=\omega^{S0}_r(0)(1-(T/T_c)^4)^{1/4}$ (see supplementary) yields $\omega^{S0}_r(0)/2\pi=10.48$~GHz.
Figure~\ref{Exp1}d shows the dependence of the coupling ratio $g/\omega^{S0}_r$ on temperature.
It shows that $g/\omega^{S0}_r$ remains above 0.5 at the entire temperature range and reaches 0.58 at $T\rightarrow0$, which at the moment is the highest coupling ratio ever reported.

Another critical characteristics of the light-matter interaction irregardless the platform is the cooperativity \cite{Huebl_PRL_111_127003,Li_PRL_123_107701} $C=g^2/\Delta\omega^{S^+}_r \Delta\omega^{S^-}_r$, where $\Delta\omega^{S^+}_r$ and $\Delta\omega^{S^-}_r$ are the linewidth of the $S^+$- and $S^-$-lines at magnetic field of coupling.
Fig.~\ref{Exp1}d shows that $C$ exceeds 9000 at $T<4$~K.
This cooperativity exceeds values for metallic-ferromagnet-based hybrids \cite{Li_PRL_123_107701,Hou_PRL_123_107702,Golovchanskiy_Sci} by more than two orders of magnitude, and becomes comparable with values for early dielectric-ferrimagnet-based hybrids \cite{Huebl_PRL_111_127003,Zhang_PRL_113_156401}.
Technically, cooperativity is the only parameter of our system, which has better alternatives \cite{Bourhill_PRB_93_144420}.

At last, the $A^2$ interaction term in Eq.~\ref{Int} should be discussed.
In general, $A^2$ term appears from the expansion of the minimal coupling Hamiltonian of a charge carries that moves in magnetic field $(\hat{p}-q\hat{A})$, where $\hat{p}$ is the momentum of a particle, $q$ is its charge, and $\hat{A}$ is the vector potential of electromagnetic field \cite{Kockum_NatPhysRev_1_19,Nataf_NatCom_1_72}.
The diamagnetic term appears naturally for plasmon-polaritons \cite{Mueller_Nat_583_780,Baranov_NatComm_11_2715,NanoLett_17_6340,Hopfield_PR_112_1555}, which considers interaction of electromagnetic field with charge oscillations (plasmons).
Its presence in our magnonic system manifests plasmonic contribution of superconducting charge carriers (cooper pairs) moving in alternating Meissner magnetic fields of the Swihart resonator. 
Its presence implies that the observed polaritons should be refereed to as the plasmon-magnon polaritons.


Summarizing, in this work we demonstrate realization of the ultra-strong photon-to-magnon coupling in thin film hetero-structures with high values of the coupling strength, single-spin coupling strength, coupling constant and cooperativity.
With the achieved coupling ratio $g/\omega\approx0.6\sim 1$ the system approaches to the deep-strong coupling regime, where the spectrum clearly evidences contribution of the diamagnetic $A^2$ interaction term in the Hamiltonian of the system.
The contribution of the $A^2$ manifests observation of a different hybrid polariton quasi-particle, namely, the plasmon-magnon polariton.
A simple increase in thickness $d_F$ by a factor of 3 should provide the coupling ratio beyond unity.  

The further opportunities for engineering of the coupling strength are rather straightforward: one can consider a microwave resonator with even smaller phase velocity that is fabricated using superconducting materials with higher magnetic penetration depth an dielectric materials with higher dielectric constant.
Using magnetic materials with lower losses, including YIG or Co$_{0.25}$Fe$_{0.75}$ will further enhance the cooperativity.
In addition, application of superconducting materials with high kinetic inductance, i.g., granular aluminium or niobium nitride, should introduce non-linearity into the system.


The Authors acknowledge Prof. Martin Weides for fruitful discussions.
This work was supported by the Ministry of Science and Higher Education of the Russian Federation, by the Russian Science Foundation, and the Russian Foundation for Basic Research.
M. Yu. K. acknowledges the support by the Interdisciplinary Scientific and Educational School of Moscow State University “Photonic and Quantum Technologies. Digital Medicine” and by the plan of scientific research of the SINP MSU.

\bibliographystyle{apsrev}
\bibliography{A_Bib_SIFS}

\end{document}